\input head
\begin{document}
\begin{titlepage}
\begin{flushright}
Z\"urich University Preprint\\
ZU-TH 27/93
\end{flushright}
\vfill
\begin{center}
{\large\bf GRAVITATIONAL MICROLENSING BY
THE DARK HALO OF M31$^{\star}$}\\
\vskip 1.0cm
{\bf Philippe~Jetzer\footnote{Supported by the Swiss National
Science Foundation}}\\
\vskip 1.0cm
Institute of Theoretical Physics,

University of Z\"urich, Winterthurerstrasse 190,\\
CH-8057 Z\"urich, Switzerland
\end{center}
\vfill
\begin{center}
Abstract
\end{center}
\baselineskip=12pt
\begin{quote}
It has been shown by Paczy\'nski that gravitational
microlensing is potentially a useful method for detecting
the dark constituents of the halo of our galaxy, if
their mass lies in the approximate domain $10^{-6}<M/M_\odot<10^{-1}$.

Microlensing observations now im progress monitor several millions of
stars in the Large Magellanic Cloud and in the Galactic Bulge.
Here I discuss the main features of the microlensing
events: in particular their rates and probability.
\end{quote}

\vfill
\begin{center}
September 1993
\end{center}
$^{\star}$ To appear in the proceedings of
the EPS conference, Marseille (France), July 1993.\\
\end{titlepage}

\baselineskip=13pt

\noindent{\bf 1. INTRODUCTION}\\

The large amounts of unobserved dark matter may
be gathered in galactic halos in a
yet-undetected form, such as ``Massive Halo Objects'' (MHOs).
Except for super-massive black holes
($M_{BH}>100M_\odot$), MHOs,
if made of ordinary matter, should be bodies made of the lightest
primordial elements: Hydrogen and Helium. In the narrow range
$0.1>M/M_\odot>0.08$ they would be hard-to-spot
white dwarfs. Lighter MHOs,
below the nuclear-ignition threshold, would be ``brown
dwarfs'', Jupiter like bodies.
The possible origin of small hydrogenoid
planets, by fragmentation or by some other mechanism, is at present
not understood.  It has been pointed out that the mass distribution
of the MHOs, normalized to the dark halo mass density, could be
a smooth continuation of the known initial mass function
of ordinary stars \cite{kn:Derujula1}.
The ambient radiation, or their own body heat, would make
sufficiently small objects made of H and He evaporate rapidly.
The condition that the rate of evaporation of such a hydrogenoid sphere be
insufficient to halve its mass in a billion years leads to the
following lower limit on their mass \cite{kn:Derujula1}:
$M > 10^{-7} M_{\odot}
(T_S /30 K)^{3/2}$ ($T_S$ being their surface
temperature and $\rho \sim 1~ g~ cm^{-3}$ their average density).

Paczy\'nski \cite{kn:Paczynski}
has shown how to
detect the MHOs if their mass lies in the range $10^{-6}
< M/M_{\odot} < 10^{-1}$ by means of the gravitational lensing effect.
Observations  which are now in progress by a French
collaboration
\cite{kn:Aubourg}
using the ESO
telescopes and  by a American-Australian collaboration
\cite{kn:Griest}
using the Mt. Stromlo Observatory
monitor several millions
stars in the Large Magellanic Cloud (LMC), whereas an American-Polish
collaboration working at Las Campanas Observatory operate on stars
located in the Galactic Bulge \cite{kn:Udalski}.\\

\noindent{\bf 2. MICROLENSING PROBABILITY}\\

When a
MHO of mass $M$ is sufficiently close to the line between us and the
star, the light from the source suffers a gravitational
deflection. The deflection angle is usually so small that we do not see
two images but rather a magnification  of the original star brightness.
This magnification, at its maximum, is given by
\begin{equation}
A_{max}=\frac{u^2+2}{u(u^2+4)^{1/2}}~ . \label{eq:bb}
\end{equation}
Here $u=d/R_E$ ($d$ is the distance of the MHO from the line of sight)
and the Einstein radius $R_E$ is defined as:
\begin{equation}
R_E^2=\frac{4GMD}{c^2}x(1-x)
\end{equation}
with $x=s/D$, and
where $D$ and $s$ are the distance between the source, respectively the MHO and
the observer.

An important quantity is the optical depth $\tau_{opt}$
to gravitational microlensing defined as
\begin{equation}
\tau_{opt}=\int_0^1 dx \frac{4\pi G}{c^2}\rho(x) D^2 x(1-x)
\label{eq:za}
\end{equation}
with $\rho(x)$ the mass density of microlensing matter at distance
$s=xD$ from us. $\tau_{opt}$ is the probability
that a source is found within a radius $R_E$ of some MHO and thus has a
magnification that is larger
than $A_{max}= 1.34$ ($d \leq R_E$).
For a spherical halo, $\tau_{opt}=0.7~ 10^{-6}$
for the LMC and $\tau_{opt}=10^{-6}$ for the SMC.

For globular clusters we get tipically
$\tau_{opt}\approx 4~ 10^{-8}$ (for instance:
NGC 288 $\tau_{opt}=6.3~ 10^{-8}$;
NGC 4833 $\tau_{opt}=5~ 10^{-8}$ and NGC 3201
$\tau_{opt}=2.8~ 10^{-8}$); compared to the
LMC or SMC this means a value $\approx$ 20 times smaller for the optical
depth $\tau_{opt}$.
However, should the planned observations of the
LMC show a lot of microlensing events, then the monitoring of
globular clusters could give, although with less data, useful
informations on the spatial distribution of the MHOs in the halo.

The magnification of the brightness of a star by a MHO is a time-dependent
effect. If both the source and the observer are at rest, the light
curve as function of time is obtained by inserting
\begin{equation}
u(t)=\frac{(d^2+v^2_Tt^2)^{1/2}}{R_E}
\end{equation}
into
eq.(\ref{eq:bb}), where $v_T$ is the transverse velocity of the MHO,
which can be inferred from the measured rotation curves ($v_T \approx 200~
km/s$). The
achromaticity, symmetry and uniqueness of the signal are distinctive
features that will allow to discriminate a microlensing event from
background events such as variable stars.\\

\noindent{\bf 3. MICROLENSING RATES}\\

Microlensing rates depend on the mass and velocity distribution of
MHOs \cite{kn:Derujula2,kn:Griest1}.
In order to quantify the expected number of events, $N_{ev}$,
for the LMC (D=50 kpc) it is convenient
to take as an example a delta function distribution for the mass of the MHOs.
For an experiment monitoring $N_{\star}$ stars during an
observation time $t_{obs}$ the total number of events with a
magnification $A \geq A_{min}$ or $u \leq u_{max}$ is given by:
\begin{equation}
N_{ev}= N_{\star}~ t_{obs}~u_{max}~ 4
\times 10^{-13}~\left( \frac{v_H}{210~km/s}\right)
 \left(\frac{1}{\sqrt{D/kpc}}\right)
 \left( \frac{\rho_0}{0.3~GeV/cm^3}\right)
\frac{1}{\sqrt{M/M_{\odot}}}~ .
\label{eq:tb}
\end{equation}
($\rho_0$ is the local dark mass density in the solar neighbourhood.)
In the following Table we show some values of $N_{ev}$ for the LMC,
taking
$t_{obs}=10^7$ sec ($\sim$ 4 Months), $N_{\star}=10^6$ stars and
$A_{min} = 1.34$.

\begin{center}
\begin{tabular}{|c|c|c|c|c|}\hline
MHO mass in units of $M_{\odot}$ & Mean $R_E$ in km & Mean microlensing time &
$N_{ev}$ \\
\hline
$10^{-2}$ & $10^8$ & 9 days & 5  \\
$10^{-4}$ & $10^7$ & 1 day & 55  \\
$10^{-6}$ & $10^6$ & 2 hours & 554  \\
\hline
\end{tabular}
\end{center}

Gravitational microlensing could also be useful for detecting MHOs in
the halo of nearby galaxies
\cite{kn:Crotts,kn:Baillon}
such as M31 or M33. In fact, it turns out
that the massive dark halo of M31 has an optical depth to microlensing
which is of about the same order of magnitude as that of our own
galaxy
\cite{kn:Crotts,kn:Jetzer}
$\sim 10^{-6}$. Moreover, an experiment monitoring stars in
M31 would be sensitive to both MHOs in our halo and in the one of M31.
One can also compute the microlensing rate
\cite{kn:Jetzer}
for MHOs in the halo of M31, for which, by taking a delta distribution
for their mass, we get
\begin{equation}
N_{ev}^a= N_{\star}~t_{obs}~u_{max}~1.8 \times 10^{-12}
\left(\frac{v_H}{210~km/s} \right)
\left(\frac{1}{\sqrt{D/kpc}}\right)
\left(\frac{\rho(0)}{1~Gev/cm^3} \right)
\frac{1}{\sqrt{M/M_{\odot}}}~. \label{eq:tc}
\end{equation}
($\rho(0)$ is the central density of the dark matter.)
In the following Table we show some values of $N^a_{ev}$ due to MHOs in the
halo of M31 with $t_{obs}= 10^7$ sec and $N_{\star}=10^6$ stars. In the
last column we give the corresponding number of events due to MHOs in our
own halo. The mean microlensing time is about the same for both types of
events.

\begin{center}
\begin{tabular}{|c|c|c|c|c|}\hline
MHO mass in units of $M_{\odot}$ & Mean $R_E$ in km & Mean microlensing time &
$N^a_{ev}$ & $N_{ev}$\\
\hline
$10^{-1}$ & $7\times 10^8$ & 38 days & 2 & 1 \\
$10^{-2}$ & $2\times 10^8$ & 12 days & 7 & 4 \\
$10^{-4}$ & $2\times 10^7$ & 30 hours & 70 & 43 \\
$10^{-6}$ & $2\times 10^6$ & 3 hours & 700 & 430 \\
\hline
\end{tabular}
\end{center}

$N_{ev}$ is almost by a factor of two bigger than $N_{ev}$. Of course these
numbers should be taken as an estimate, since they depend on the details
of the model one adopts for the distribution of the dark matter in the halo.\\

\begin{figcap}
\item Plot of $-\Delta m=2.5~log A(t)$ as a function of time (in units of
$10^5$ sec) for a pointlike source at the distance of the LMC. The MHO is at
a distance of 25 kpc and has a transverse velocity of 200 km/s and a mass
of $10^{-2} M_{\odot}$. The higher curve, that serves to define the event
duration T, corresponds to $d = 1/2~ R_E$, whereas the lower one has
$d = R_E$.
\end{figcap}
\end{document}